\documentclass[twocolumn,letter]{jpsj3}
\usepackage{graphicx}
\usepackage{dcolumn}
\usepackage{bm}
\usepackage{ulem}
\usepackage{braket}
\usepackage{amsmath}
\usepackage{color}

\title{Stability of Skyrmion Crystal Phase in Centrosymmetric Distorted Triangular-Lattice Antiferromagnets}

\author{Satoru Hayami}
\inst{
Department of Applied Physics, The University of Tokyo, Tokyo 113-8656, Japan
 }

\abst{
The stability of a magnetic skyrmion crystal in a centrosymmetric orthorhombic lattice system is numerically investigated. 
By performing the simulated annealing for an effective spin model with the momentum-resolved interaction on a uniaxially distorted triangular lattice, we find that the skyrmion crystal is robustly realized when dominant interaction channels consist of the double-$Q$ structure in momentum space. 
Moreover, we show that an anti-type skyrmion crystal with the opposite skyrmion number is obtained by tuning the sign of the bond-dependent anisotropic exchange interaction that originates from the relativistic spin-orbit coupling. 
}

\begin{document}
\maketitle

Magnetic frustration that originates from competing interactions in magnetic systems has brought about intriguing antiferromagnetic spin configurations and their associated response and transport phenomena~\cite{diep2004frustrated, lacroix2011introduction, Nisoli_RevModPhys.85.1473, batista2016frustration}. 
A canonical model of magnetic frustration is the Heisenberg model with the nearest-neighbor ferromagnetic interaction and further-neighbor antiferromagnetic interaction, which leads to coplanar spiral spin states with different magnetic periods according to the ratio of the exchange interactions~\cite{tamura2011first}. 
Such a competing interaction becomes a source of noncoplanar multiple-$Q$ spin states consisting of multiple spiral waves including a skyrmion crystal (SkX) by additionally taking into account the effect of an external magnetic field, single-ion anisotropy, non-magnetic impurity, and thermal/quantum fluctuations even in centrosymmetric magnets~\cite{Binz_PhysRevLett.96.207202,Binz_PhysRevB.74.214408,Okubo_PhysRevLett.108.017206,nagaosa2013topological,Kamiya_PhysRevX.4.011023,leonov2015multiply,Lin_PhysRevB.93.064430,Hayami_PhysRevB.93.184413,Hayami_PhysRevB.94.174420,Lin_PhysRevLett.120.077202,Shimokawa_PhysRevB.100.224404,Balla_PhysRevResearch.2.043278,Hayami_PhysRevB.103.224418,Pohle_PhysRevB.104.024426,hayami2022skyrmion,Aoyama_PhysRevB.105.L100407}. 
Furthermore, recent studies have revealed that a new type of magnetic frustration in momentum space that arises from the itinerant nature of electrons, which is termed itinerant frustration~\cite{hayami2021topological}, also gives rise to various multiple-$Q$ states~\cite{Martin_PhysRevLett.101.156402, Akagi_PhysRevLett.108.096401, Hayami_PhysRevB.90.060402, Hayami_PhysRevB.94.024424,Ozawa_doi:10.7566/JPSJ.85.103703, Ozawa_PhysRevLett.118.147205, Hayami_PhysRevB.95.224424, Hayami_PhysRevB.99.094420, Wang_PhysRevLett.124.207201,hayami2021phase,eto2022low}.

Although these frustration-based mechanisms are ubiquitous irrespective of lattice symmetry, they have been mainly studied in high-symmetry lattice systems to have the $n$-fold rotation axis ($n=3,4,6$). 
This is because a degeneracy between different single-$Q$ spiral states connected by lattice symmetry tends to be the origin of multiple-$Q$ states by forming their superposition. 
Indeed, the multiple-$Q$ states have been observed in centrosymmetric high-symmetry lattice systems with hexagonal symmetry as Y$_3$Co$_8$Sn$_4$~\cite{takagi2018multiple}, Gd$_2$PdSi$_3$~\cite{kurumaji2019skyrmion,Kumar_PhysRevB.101.144440,spachmann2021magnetoelastic,paddison2022magnetic}, and Gd$_3$Ru$_4$Al$_{12}$~\cite{chandragiri2016magnetic,Nakamura_PhysRevB.98.054410,hirschberger2019skyrmion,Hirschberger_10.1088/1367-2630/abdef9}, tetragonal symmetry as GdRu$_2$Si$_2$~\cite{khanh2020nanometric,Yasui2020imaging,khanh2022zoology}, EuAl$_4$~\cite{Shang_PhysRevB.103.L020405,kaneko2021charge,Zhu_PhysRevB.105.014423,takagi2022square,meier2022thermodynamic}, and CeAuSb$_2$~\cite{Marcus_PhysRevLett.120.097201,Seo_PhysRevX.10.011035,seo2021spin}, and  cubic symmetry as SrFeO$_3$~\cite{Ishiwata_PhysRevB.84.054427,yambe2020double,Ishiwata_PhysRevB.101.134406,Rogge_PhysRevMaterials.3.084404,Onose_PhysRevMaterials.4.114420} and MnSc$_2$S$_4$~\cite{Gao2016Spiral,gao2020fractional}. 
Another feature of the frustration-based mechanisms is to keep degeneracy for the spin texture due to the presence of the global spin rotation symmetry, which leads to a coexisting state of skyrmions and anti-skyrmions~\cite{Okubo_PhysRevLett.108.017206}, a replica-symmetry-breaking SkX state~\cite{Mitsumoto_PhysRevB.104.184432}, and their dynamics~\cite{leonov2015multiply,zhang2015magnetic, Lin_PhysRevB.93.064430,zhang2020skyrmion, Yao_PhysRevB.105.014444}. 

With such situations in mind, we aim at illuminating the stability of the SkX in centrosymmetric low-symmetry lattice systems without $n$-fold rotation axis ($n=3,4,6$) in order to further stimulate an exploration of multiple-$Q$ states in both theoretical and experimental studies. 
For that purpose, we consider a uniaxially distorted triangular lattice belonging to the orthorhombic point group with twofold rotation symmetry, which has often been studied in noncentrosymmetric systems where the modulated Dzyaloshinskii-Moriya interaction becomes important~\cite{Bogdanov_PhysRevLett.87.037203,shibata2015large, Wang_PhysRevB.97.024429, Osorio_PhysRevB.100.220404, Hog_PhysRevB.104.024402}. 
As another aspect, we focus on the degeneracy lifting between the SkX and anti-SkX with opposite-sign skyrmion numbers by considering the magnetic anisotropy arising from discrete rotational symmetry. 
For example, the energy in the SkX is usually lower than that in the anti-SkX even in the centrosymmetric lattice structure in the presence of the sixfold-symmetric magnetic anisotropy arsing from spin-orbit coupling in the hexagonal systems~\cite{amoroso2020spontaneous, yambe2021skyrmion, Hayami_PhysRevB.103.054422, amoroso2021tuning, Utesov_PhysRevB.105.054435, yambe2022effective}. 
The creation of the anti-SkX in such a situation has not been elucidated thus far.

To obtain an insight into the above issues, we analyze an effective spin model with both isotropic and anisotropic exchange interactions on the distorted triangular lattice. 
By constructing the magnetic phase diagrams while varying the ratio of the interactions for different bond directions through the simulated annealing, we find that the Bloch SkX is robustly stable in the presence of the distinct double-$Q$ peak structure in the momentum-resolved interaction. 
Meanwhile, we show that the SkX is replaced with the single-$Q$ spiral state when the momentum-resolved interaction exhibits the single-$Q$ peak structure. 
Furthermore, we find that the anti-SkX can be realized by taking a different ratio to the isotropic and anisotropic exchange interactions. 
Our results provide an important essence to inducing the SkX and anti-SkX in the orthorhombic systems.

\begin{figure}[t!]
\begin{center}
\includegraphics[width=1.0\hsize]{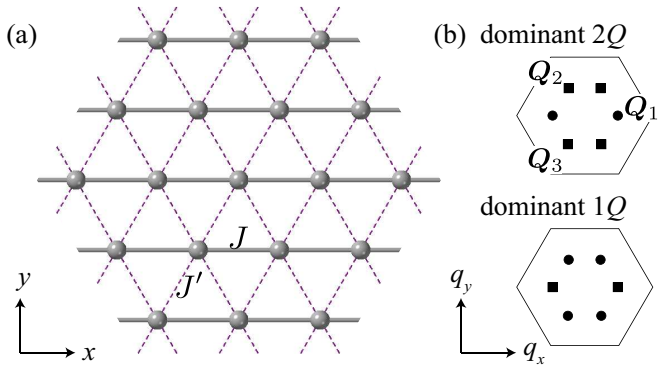} 
\caption{
\label{fig:fig1}
(Color online) 
(a) Distorted triangular-lattice structure. $J$ and $J'$ represent the intra- and inter-chain exchange interactions, respectively.  
(b) Two cases of the triple-$Q$ peak structures in the momentum-resolved exchange interaction at $\bm{Q}_1$, $\bm{Q}_2$, and $\bm{Q}_3$. 
The squares and circles represent the dominant and sub-dominant peak positions, respectively. 
}
\end{center}
\end{figure}

We consider the uniaxially distorted triangular lattice in Fig.~\ref{fig:fig1}(a), where the one-dimensional chains along the $x$ direction are stacked along the $y$ direction. 
We take the distances of the nearest-neighbor spins along the intra- and inter-chain directions as unity for simplicity; the difference of the bonds is described by the different magnitude of the exchange interactions denoted as $J$ and $J'$ in Fig.~\ref{fig:fig1}(a). 
When setting $J=J'$, the system reduces to the triangular-lattice system. 

We analyze an effective spin model with the momentum-resolved interaction, which is given by 
\begin{align}
\mathcal{H} = -2\sum_{\nu}\sum_{\alpha,\beta}J_{\bm{Q}_\nu} I^{\alpha \beta}_{\bm{Q}_{\nu}} S^\alpha_{\bm{Q}_{\nu}} S^\beta_{-\bm{Q}_{\nu}} 
 -H \sum_{i} S^z_i, 
\label{eq: Ham}
\end{align}
where $\bm{S}_i=(S_i^x, S_i^y, S_i^z)$ is the classical localized spin at site $i$ and $S^\alpha_{\bm{Q}_\nu}$ is the $\bm{Q}_\nu$ ($\nu$ is the index for the wave vector) and $\alpha$ ($\alpha=x,y,z$) components of the Fourier transform of $S^\alpha_i$; we fix $|\bm{S}_i|=1$. 
The first term represents the momentum-resolved exchange interaction with the coupling constant $J_{\bm{Q}_\nu} $, where $I^{\alpha\beta}_{\bm{Q}_{\nu}}$ represents the form factor for given $\bm{Q}_\nu$, and $\alpha, \beta$. 
We adopt a bond-dependent anisotropic exchange interaction in addition to the isotropic Heisenberg-type interaction; $I^{\alpha\beta}_{\bm{Q}_{\nu}}$ is divided into $I^{\alpha\beta}_{\bm{Q}_{\nu}}=\delta^{\alpha\beta}+\tilde{I}^{\alpha\beta}_{\bm{Q}_{\nu}}$, where $\delta^{\alpha\beta}$ is the Kronecker delta and $\tilde{I}^{\alpha\beta}_{\bm{Q}_{\nu}}$ depends on the direction of $\bm{Q}_{\nu}$ while satifying the symmetry of the lattice structure. 
This bond-dependent anisotropic interaction originates from relativistic spin-orbit coupling, which not only lifts the degeneracy between the SkX and anti-SkX but also leads to various vortex spin textures in the hexagonal systems~\cite{Michael_PhysRevB.91.155135,Rousochatzakis2016,Li_PhysRevResearch.1.013002,amoroso2020spontaneous,Hayami_PhysRevB.104.094425} and the tetragonal systems~\cite{Hayami_PhysRevLett.121.137202,Hayami_doi:10.7566/JPSJ.89.103702,Hayami_PhysRevB.103.024439,Utesov_PhysRevB.103.064414,Wang_PhysRevB.103.104408,Kato_PhysRevB.104.224405,Hayami_PhysRevB.105.104428}. 
We ignore the other anisotropic exchange interactions and single-ion anisotropy for simplicity. 
The second term in Eq.~(\ref{eq: Ham}) represents the Zeeman coupling under an external magnetic field along the $z$ direction. 

The momentum-resolved interaction $J_{\bm{Q}_\nu}$ is a consequence of the interaction in real space. 
There are mainly two origins: One is the short-range interactions in frustrated magnets and the other is the long-range ones in itinerant magnets. 
For the former, the competition between the nearest-neighbor interaction and the further-neighbor interaction plays an important role, while the Fermi surface instability of itinerant electrons inducing the Ruderman-Kittel-Kasuya-Yosida interaction~\cite{Ruderman, Kasuya, Yosida1957} is important for the latter. 
In both cases, the wave vector dependence of $J_{\bm{Q}_\nu}$ is obtained by performing the Fourier transform of the real-space interaction. 

In the following, we analyze the ground state of the model in Eq.~(\ref{eq: Ham}) by extracting the dominant $\bm{Q}_\nu$ contribution and by dropping off the irrelevant wave vector contribution. 
In order to compare the result of the triangular-lattice case, we consider the interaction at triple-$Q$ wave vectors connected by the threefold rotational symmetry: $\bm{Q}_1=(Q,0)$, $\bm{Q}_2=(-Q/2,\sqrt{3}Q/2)$, and $\bm{Q}_3=(-Q/2,-\sqrt{3}Q/2)$ with $Q=\pi/3$.  
In the case of the isotropic triangular lattice, the interactions at $\bm{Q}_\nu$ are equivalent, i.e., $J_{\bm{Q}_1}=J_{\bm{Q}_2}=J_{\bm{Q}_3}$.
When introducing the uniaxial distortion of the triangular lattice, its threefold rotational symmetry is broken; the interactions at $J_{\bm{Q}_\nu}$ are no longer equivalent and $\bm{Q}_\nu$ are not symmetry-related.
Under these circumstances, we incorporate the effect of the uniaxial deformation by setting the different interaction parameters for $J_{\bm{Q}_1} \equiv \alpha_{1Q} \tilde{J}$ and $J_{\bm{Q}_2}=J_{\bm{Q}_3}=\alpha_{2Q}\tilde{J}$ [$\tilde{J}=1$ is the energy unit of the model in Eq.~(\ref{eq: Ham})] and neglect the change of the dominant wave-vector positions for simplicity.
Similar qualitative results are expected even for the different wave-vector positions once the relation $\bm{Q}_1+\bm{Q}_2+\bm{Q}_3=\bm{0}$ is satisfied. 
In addition, we treat $\alpha_{1Q}$ and $\alpha_{2Q}$ as phenomenological parameters, although they are related to the interaction in real space. 
It is noted that the interactions at $\bm{Q}_2$ and $\bm{Q}_3$ are equivalent, since $\bm{Q}_2$ and $\bm{Q}_3$ are connected by vertical mirror symmetry. 

The simplification of the interaction is justified when considering the low-temperature spin configuration, where its energy is optimized by the dominant ordering wave vectors. 
Similar attempts based on similar effective spin models have been studied in both frustrated~\cite{leonov2015multiply,Hayami_PhysRevB.103.224418,Hayami_PhysRevB.105.014408,hayami2022multiple,hayami2022skyrmion} and itinerant magnets~\cite{Hayami_PhysRevB.95.224424,hayami2020multiple,Wang_PhysRevLett.124.207201,Hayami_PhysRevB.103.054422,yambe2021skyrmion,hayami2021field,yambe2022effective,Kato_PhysRevB.105.174413}, where the instability toward the SkX has been found.  
Besides, the stability of the multiple-$Q$ states other than the SkX by taking the different amplitudes of the interactions at relevant $\bm{Q}_\nu$ has been investigated in the cubic-tetragonal system~\cite{Shimizu_PhysRevB.103.054427}. 
Hereafter, we consider two cases: One is the dominant double-$Q$ structure by setting $\alpha_{1Q}<1$ and $\alpha_{2Q}=1$ in the upper panel of Fig.~\ref{fig:fig1}(b) and the other is the dominant single-$Q$ structure by setting $\alpha_{1Q}=1$ and $\alpha_{2Q}<1$ in the lower panel of Fig.~\ref{fig:fig1}(b)

Under the assumption, the nonzero form factor is given by $
-\tilde{I}^{xx}_{\bm{Q}_{1}}=\tilde{I}^{yy}_{\bm{Q}_{1}}=2\tilde{I}^{xx}_{\bm{Q}_{2}}=-2\tilde{I}^{yy}_{\bm{Q}_{2}}=2\tilde{I}^{xy}_{\bm{Q}_{2}}/\sqrt{3}=2\tilde{I}^{yx}_{\bm{Q}_{2}}/\sqrt{3}=2\tilde{I}^{xx}_{\bm{Q}_{3}}=-2\tilde{I}^{yy}_{\bm{Q}_{3}}=-2\tilde{I}^{xy}_{\bm{Q}_{3}}/\sqrt{3}=-2\tilde{I}^{yx}_{\bm{Q}_{3}}/\sqrt{3} \equiv I^{\rm A}$~\cite{Hayami_PhysRevB.103.054422,Hirschberger_10.1088/1367-2630/abdef9}. 
We set $I^{\rm A}=0.1$ to fix the spiral plane so as to form the proper-screw spiral or the Bloch SkX. 
In the case of the triangular-lattice model, i.e., $\alpha_{1Q}=\alpha_{2Q}=1$, the model exhibits the SkX for nonzero $I^{\rm A}$, as detailed below~\cite{Hayami_PhysRevB.103.054422}. 

The stability of the SkX under $\alpha_{1Q}<1$ or $\alpha_{2Q}<1$ is investigated by performing the simulated annealing based on the Metropolis local updates for $\bm{S}_i$ in real space~\cite{metropolis1953equation}. 
In each simulation, the temperature is gradually reduced from high temperatures $T_0 = 1$-$10$ to the lowest temperature $T=0.001$ with the rate of $T_{n+1} = \alpha T_n$ at $\alpha = 0.99999$-$0.999999$ ($T_n$ is the temperature in the $n$th step). 
At the final temperature, $10^5$-$10^6$ Monte Carlo sweeps are carried out after equilibration. 
We also start the simulations from the spin configurations in the vicinity of the phase boundaries.
The numerical simulations are performed for the system size with $N=48^2$.

\begin{figure}[t!]
\begin{center}
\includegraphics[width=1.0\hsize]{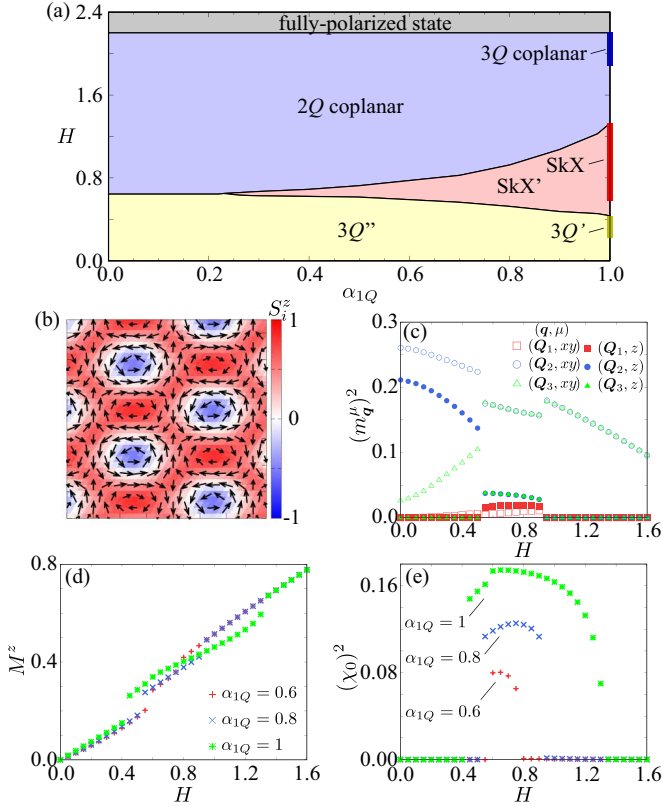} 
\caption{
\label{fig:fig2}
(Color online) 
(a) Phase diagram in the plane of $\alpha_{1Q}$-$H$ at $I^{\rm A}=0.1$ and $\alpha_{2Q}=1$. 
(b) Snapshots of the real-space spin configuration in the SkX' phase at $\alpha_{1Q}=0.6$ and $H=0.65$. 
The arrows and color represent the $xy$ and $z$ components of the spin moment, respectively. 
(c) $H$ dependences of $(m^\mu_{\bm{q}})^2$ for $\mu=xy, z$ and $\bm{q}=\bm{Q}_1, \bm{Q}_2, \bm{Q}_3$ at $\alpha_{1Q}=0.8$. 
(d,e) $H$ dependences of (d) $M^z$ and (e) $(\chi_0)^2$ for $\alpha_{1Q}=0.6$, $0.8$, and $1$. 
}
\end{center}
\end{figure}

Figure~\ref{fig:fig2}(a) shows the phase diagram while changing $\alpha_{1Q}$ and $H$ for fixed $\alpha_{2Q}$, which corresponds to the dominant double-$Q$ structure in the upper panel of Fig.~\ref{fig:fig1}(b). 
In the phase diagram, there are six phases with different spin configurations in addition to the fully-polarized state for $H \gtrsim 2.2$; $2Q$ and $3Q$ stand for the double-$Q$ and triple-$Q$ spin textures, respectively, where ' and '' for $Q$ and SkX mean the different amplitudes of the spin structure factor to satisfy $|\bm{m}_{\bm{Q}_1}| \neq |\bm{m}_{\bm{Q}_2}| = |\bm{m}_{\bm{Q}_3}|$ for $3Q$' and SkX' and $|\bm{m}_{\bm{Q}_1}| \neq |\bm{m}_{\bm{Q}_2}| \neq |\bm{m}_{\bm{Q}_3}|$ for $3Q$''. 
Here, $\bm{m}_{\bm{Q}_\nu}$ is defined as $m^\alpha_{\bm{q}}=\sqrt{S^{\alpha\alpha}(\bm{q})/N}$, where $S^{\alpha\alpha}(\bm{q})=(1/N)\sum_{i,j}S^{\alpha}_i S^{\alpha}_j e^{i\bm{q}\cdot (\bm{r}_i -\bm{r}_j)}$ is the spin structure factor for $\alpha=x,y,z$ and $\bm{q}=\bm{Q}_1, \bm{Q}_2, \bm{Q}_3$. 
We also define $(m^{xy}_{\bm{Q}_\nu})^2=(m^{x}_{\bm{Q}_\nu})^2+(m^{y}_{\bm{Q}_\nu})^2$. 
We show the behavior of $(m^{\mu}_{\bm{Q}_\nu})^2$ ($\mu=xy, z$) against $H$ at $\alpha_{1Q}=0.8$ is shown in Fig.~\ref{fig:fig2}(c). 
In addition, we plot the $H$ dependences of the uniform magnetization $M^z=(1/N)\sum_i S_i^z$ and the scalar chirality $(\chi_0)^2=[(1/N)\sum_{\bm{R}}\bm{S}_i \cdot (\bm{S}_j \times \bm{S}_k)]^2$ ($\bm{R}$ represents the position vector at the triangle $\bm{R}$ consisting of site $i$, $j$, and $k$) for several $\alpha_{1Q}$ in Figs.~\ref{fig:fig2}(d) and \ref{fig:fig2}(e), respectively. 

For $\alpha_{1Q}=1$, two types of SkXs denoted as SkX and SkX' appear in the intermediate field, which is consistent with the previous result~\cite{Hayami_PhysRevB.103.054422}.
Although both SkX phases are characterized by the triple-$Q$ spin configurations, the SkX exhibits equal intensity as $|\bm{m}_{\bm{Q}_1}| = |\bm{m}_{\bm{Q}_2}| = |\bm{m}_{\bm{Q}_3}|$, while the SkX' exhibits different intensities as $|\bm{m}_{\bm{Q}_1}| \neq |\bm{m}_{\bm{Q}_2}| = |\bm{m}_{\bm{Q}_3}|$. 
The skyrmion number in both states is quantized as $-1$; there is no anti-SkX. 
In the low(high)-field region, the 3$Q$' and 3$Q$'' (2$Q$ coplanar and 3$Q$ coplanar~\cite{comment_coplanar}) states are stabilized; the 3$Q$' and 3$Q$'' states are mainly described by the single-$Q$ proper-screw spiral state with additional $m^{xy}_{\bm{Q}_2}=m^{xy}_{\bm{Q}_3}$ and $m^{xy}_{\bm{Q}_2} \neq m^{xy}_{\bm{Q}_3}$, respectively. 
Meanwhile, the 2$Q$ coplanar and 3$Q$ coplanar states are characterized by the equivalent in-plane modulation as $m^{xy}_{\bm{Q}_2}=m^{xy}_{\bm{Q}_3}$ and $m^{xy}_{\bm{Q}_1}=m^{xy}_{\bm{Q}_2}=m^{xy}_{\bm{Q}_3}$, respectively. 
As shown in Figs.~\ref{fig:fig2}(c)-\ref{fig:fig2}(e), the spin and chirality quantities jump at the transitions to the SkX phase. 
The emergence of these multiple-$Q$ states is attributed to nonzero $I^{\rm A}$. 

When distorting the triangular lattice as $\alpha_{1Q}<1$, the SkX, 3$Q$', and 3$Q$ coplanar phases vanish owing to the symmetry lowering of the system, each of which turns into the SkX', 3$Q$'', and $2Q$ coplanar phases, respectively, as shown in Fig.~\ref{fig:fig2}(a). 
One finds that the SkX' phase is robustly stabilized for relatively small $\alpha_{1Q} \simeq 0.23$. 
The real-space spin configuration in the SkX' phase is shown in Fig.~\ref{fig:fig2}(b), where the SkX cores are located at the center of the plaquette around $S_i^z = -1$~\cite{Hayami_PhysRevResearch.3.043158} and are elongated along the $x$ direction. 
This result indicates that the spin configuration is mainly characterized by the double-$Q$ modulation along the $\bm{Q}_2$ and $\bm{Q}_3$ directions. 
Indeed, $(m^{\mu}_{\bm{Q}_1})^2$ is much smaller than $(m^{\mu}_{\bm{Q}_2})^2=(m^{\mu}_{\bm{Q}_3})^2$, as shown in Fig.~\ref{fig:fig2}(c). 
It is noted that the ordering vector $\bm{Q}_1$ is regarded as the higher-harmonic vectors of $\bm{Q}_2$ and $\bm{Q}_3$ due to $\bm{Q}_2+\bm{Q}_3= -\bm{Q}_1$. 
In this context, the interaction at $\bm{Q}_1$ assists the stabilization of the square-like SkX, which has been recently revealed in the case of the tetragonal systems~\cite{hayami2022multiple,takagi2022square,hayami2022rectangular}. 
In other words, the system approaches the square-lattice system while decreasing $\alpha_{1Q}$. 
Since there are almost no amplitudes of $\bm{m}_{\bm{Q}_1}$ in the 3$Q$'' and 2$Q$ coplanar states, the SkX' region gradually becomes narrower while decreasing $\alpha_{1Q}$. 
Accordingly, $(\chi_0)^2$ becomes smaller while decreasing $\alpha_{1Q}$, as shown in Fig.~\ref{fig:fig2}(e).

\begin{figure}[t!]
\begin{center}
\includegraphics[width=1.0\hsize]{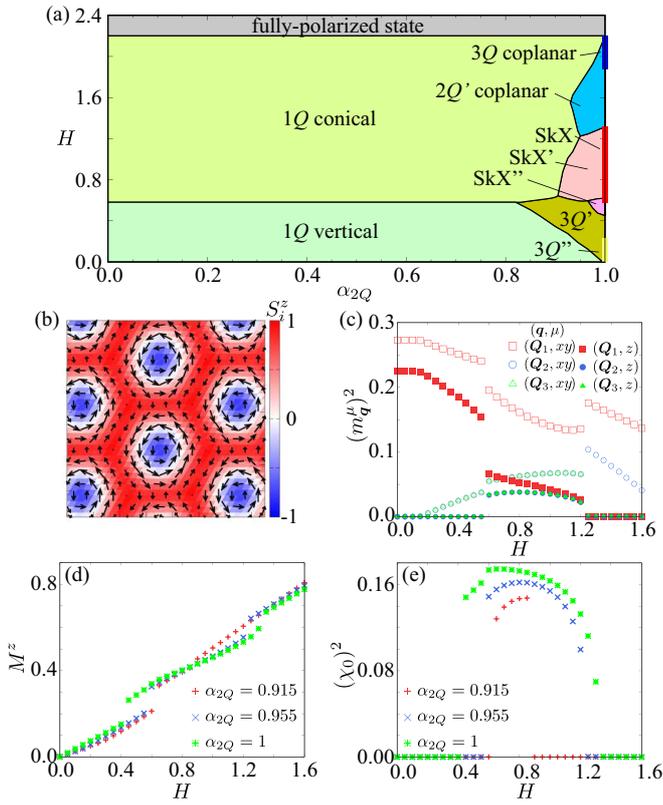} 
\caption{
\label{fig:fig3}
(Color online) 
(a) Phase diagram in the plane of $\alpha_{2Q}$-$H$ at $I^{\rm A}=0.1$ and $\alpha_{1Q}=1$. 
(b) Snapshots of the real-space spin configuration in the SkX' phase at $\alpha_{1Q}=0.955$ and $H=0.8$. 
The arrows and color represent the $xy$ and $z$ components of the spin moment, respectively. 
(c) $H$ dependences of $(m^\mu_{\bm{q}})^2$ for $\mu=xy, z$ and $\bm{q}=\bm{Q}_1, \bm{Q}_2, \bm{Q}_3$ at $\alpha_{2Q}=0.955$. 
(d,e) $H$ dependences of (d) $M^z$ and (e) $(\chi_0)^2$ for $\alpha_{2Q}=0.915$, $0.955$, and $1$. 
}
\end{center}
\end{figure}

Next, we consider the dominant single-$Q$ structure in the lower panel of Fig.~\ref{fig:fig1}(b), i.e., $\alpha_{2Q} \leq \alpha_{1Q}$. 
We show the phase diagram against $\alpha_{2Q}$ and $H$ for fixed $\alpha_{1Q}=1$ in Fig.~\ref{fig:fig3}(a), which includes additional phases, the SkX'', 2$Q$' coplanar, 1$Q$ conical, and 1$Q$ vertical spiral phases, compared to the phase diagram in Fig.~\ref{fig:fig2}(a). 
Among them, the 1$Q$ conical and 1$Q$ vertical spiral phases stabilized in the almost entire region are characterized by the spiral modulation perpendicular to the $\hat{\bm{z}}$ and $\bm{Q}_\nu$, respectively. 
The SkX'' (2$Q$' coplanar state) is identified as the spin configuration with $|\bm{m}_{\bm{Q}_1}| \neq |\bm{m}_{\bm{Q}_2}| \neq |\bm{m}_{\bm{Q}_3}|$ ($|m^{xy}_{\bm{Q}_1}| \neq |m^{xy}_{\bm{Q}_2}|$ and $m^{xy}_{\bm{Q}_3}=0$). 
Thus, there are two types of SkX phases (SkX' and SkX'') for $\alpha_{2Q}<1$. 
It is noted that both SkX phases have the skyrmion number of $-1$; there is no anti-SkX. 

As shown in Fig.~\ref{fig:fig3}(a), the SkX' rather than the SkX'' is mainly stabilized in the presence of the distortion. 
The spin configuration in the SkX' phase is characterized by the anisotropic triple-$Q$ spin configuration similar to the SkX' in Fig.~\ref{fig:fig2}(c). 
Besides, the behaviors of $M^z$ and $(\chi_0)^2$ in Figs.~\ref{fig:fig3}(d) and \ref{fig:fig3}(e) are similar to those in Figs.~\ref{fig:fig2}(d) and \ref{fig:fig2}(e). 
Meanwhile, the dominant peak is found at the $\bm{Q}_1$ component of the spin moment satisfying $(m^{\mu}_{\bm{Q}_1})^2  > (m^{\mu}_{\bm{Q}_2})^2=(m^{\mu}_{\bm{Q}_3})^2$ owing to $\alpha_{2Q}<\alpha_{1Q}$, as shown in Fig.~\ref{fig:fig3}(c).  
Accordingly, the SkX cores are elongated along the $y$ direction in Fig.~\ref{fig:fig3}(b). 
Furthermore, the stability of the SkX' phase against $\alpha_{2Q}$ is different from that against $\alpha_{1Q}$, as compared to the results in Figs.~\ref{fig:fig2}(a) and \ref{fig:fig3}(a); the region of the SkX' phase for the latter becomes narrower and is extended up to $\alpha_{2Q} \simeq 0.915$. 
This result indicates that the dominant double-$Q$ peak structure is preferred to realize the SkX spin texture rather than the dominant single-$Q$ one.

\begin{figure}[t!]
\begin{center}
\includegraphics[width=1.0\hsize]{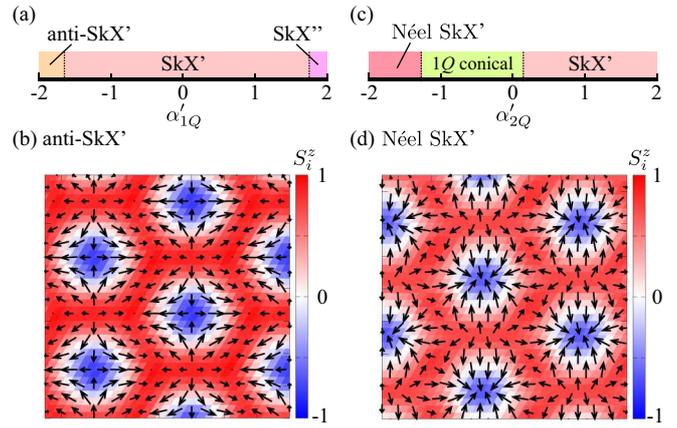} 
\caption{
\label{fig:fig4}
(Color online) 
(a) Phase diagram against $\alpha'_{1Q}$ for fixed $\alpha_{1Q}=0.95$, $\alpha_{2Q}=\alpha'_{2Q}=1$, and $H=0.7$. 
(b) Snapshots of the spin configuration in the anti-SkX' phase at $\alpha'_{1Q}=-2$. 
(c) Phase diagram against $\alpha'_{2Q}$ for fixed $\alpha_{1Q}=\alpha'_{1Q}=1$, $\alpha_{2Q}=0.95$, and $H=0.7$. 
(d) Snapshots of the spin configuration in the N\'eel-SkX' phase at $\alpha'_{2Q}=-2$. 
}
\end{center}
\end{figure}

Finally, let us discuss the possibility of the anti-SkX under the distortion. 
Although we consider the same parameters $\alpha_{1Q}$ and $\alpha_{2Q}$ for the isotropic and bond-dependent anisotropic exchange interactions, they can take different values from the orthorhombic symmetry. 
We here introduce additional parameters $\alpha'_{1Q}$ and $\alpha'_{2Q}$ for the anisotropic exchange interaction by setting $J_{\bm{Q}_1} I^{\alpha\beta}_{\bm{Q}_{1}}=J \alpha_{1Q} (\delta^{\alpha\beta}+\alpha'_{1Q}\tilde{I}^{\alpha\beta}_{\bm{Q}_{1}})$ and $J_{\bm{Q}_{2,3}} I^{\alpha\beta}_{\bm{Q}_{2,3}}=J\alpha_{2Q} (\delta^{\alpha\beta}+\alpha'_{2Q}\tilde{I}^{\alpha\beta}_{\bm{Q}_{2,3}})$. 

Figure~\ref{fig:fig4}(a) shows the phase diagram while varying $\alpha'_{1Q}$ for $\alpha_{1Q}=0.95$, $\alpha_{2Q}=\alpha'_{2Q}=1$, and $H=0.7$. 
One finds that the anti-SkX' appears for large negative $\alpha'_{1Q}$ in Fig.~\ref{fig:fig4}(a), whose real-space spin configuration is shown in Fig.~\ref{fig:fig4}(b). 
In contrast to the SkX' in Fig.~\ref{fig:fig2}(b), the spins around the core are rotated in an opposite way so as to have the skyrmion number of $+1$ in Fig.~\ref{fig:fig4}(b). 
The emergence of the anti-SkX' is owing to the negative sign of $\alpha'_{1Q}$, where the negative anisotropic exchange interaction tends to favor the out-of-plane cycloidal spiral wave instead of the proper-screw one. 
Thus, a superposition of the out-of-plane cycloidal spiral wave with $\bm{Q}_1$ and the proper-screw spiral wave with $\bm{Q}_2$ and $\bm{Q}_3$, which are favored by the anisotropic exchange interaction, is a key ingredient to give rise to the anti-SkX. 

On the other hand, the instability toward the anti-SkX' is not obtained while changing $\alpha'_{2Q}$ for $\alpha_{1Q}=\alpha'_{1Q}=1$, $\alpha_{2Q}=0.95$, and $H=0.7$, as shown in Fig.~\ref{fig:fig4}(c). 
Instead, the N\'eel SkX', which has a different helicity from the SkX' in Fig.~\ref{fig:fig3}(b), appears for large negative $\alpha'_{2Q}$. 
The spin configuration in real-space is shown in Fig.~\ref{fig:fig4}(d). 
As the N\'eel SkX' is described by a superposition of the out-of-plane cycloidal spiral wave with $\bm{Q}_1$, $\bm{Q}_2$, and $\bm{Q}_3$, there is an energy gain for the $\bm{Q}_2$ and $\bm{Q}_3$ components but an energy loss for the $\bm{Q}_1$ component. 
To avoid such an energy loss, $(m^{xy}_{\bm{Q}_{2,3}})^2$ is much larger than $(m^{xy}_{\bm{Q}_{1}})^2$ in the N\'eel SkX'; $(m^{xy}_{\bm{Q}_{1}})^2 \simeq 0.005$ and $(m^{xy}_{\bm{Q}_{2,3}})^2 \simeq 0.167$ for $\alpha'_{2Q}=-2$. 
It is noted that the SkX' phase is replaced with the 1$Q$ conical phase for small $\alpha'_{2Q}$, which implies the anisotropic exchange interaction at $\bm{Q}_2$ and $\bm{Q}_3$ plays an important role in stabilizing the SkX' in the dominant single-$Q$ case.

In summary, we have investigated the stability of the SkX under the uniaxial distortion which breaks the threefold rotational symmetry of the triangular lattice. 
Based on the simulated annealing for the effective spin model incorporating the effect of the momentum- and bond-dependent anisotropic interactions, we constructed the magnetic phase diagrams under the external magnetic field in two situations: One is the dominant double-$Q$ structure in the interaction and the other is the dominant single-$Q$ one. 
As a result, we found that the SkX is robustly stabilized in the dominant double-$Q$ structure, while it rapidly becomes unstable in the dominant single-$Q$ structure. 
In addition, we have discussed the possibility of the anti-SkX by considering the different ratios of the isotropic and anisotropic exchange interactions. 
We showed that the anti-SkX can be stabilized for large negative $\alpha'_{1Q}$ in the dominant double-$Q$ structure, while it is not realized in the dominant single-$Q$ structure. 

\begin{acknowledgments}
This research was supported by JSPS KAKENHI Grants Numbers JP19H01834, JP21H01037, JP22H04468, and by JST PRESTO (JPMJPR20L8). 
Parts of the numerical calculations were performed in the supercomputing systems in ISSP, the University of Tokyo.
\end{acknowledgments}

\bibliographystyle{JPSJ}
\bibliography{ref}

\end{document}